\documentclass[]{emulateapj}
\usepackage{apjfonts}

\usepackage{epsfig,graphicx,latexsym,amsmath,amssymb}
\usepackage{lastpage}
\usepackage{color}

\def\comment#1{}

\begin{document}

\title{A New Type of Extreme-mass-ratio Inspirals Produced by Tidal Capture of Binary Black Holes}

\author{Xian Chen\altaffilmark{1,2}}
\altaffiltext{1}{Astronomy Department, School of Physics, Peking University,
Beijing 100871, China}
\altaffiltext{2}{Kavli Institute for Astronomy and Astrophysics at 
Peking University, Beijing 100871, China}

\author{Wen-Biao Han\altaffilmark{3,4}$^\dagger$}
\altaffiltext{3}{Shanghai Astronomical Observatory, Shanghai 200030, China}
\altaffiltext{4}{School of Astronomy and Space Science, University of Chinese Academy of Sciences,  Beijing 100049, China }
\email{$^\dagger$Corresponding author, Email: wbhan@shao.ac.cn}

\date{\today}

\begin{abstract}
Extreme-mass-ratio inspiral (EMRI) is an important gravitational-wave (GW)
source and it normally consists of one stellar-mass black hole (BH) whirling
closely around a supermassive black hole (SMBH).  In this Letter, we
demonstrate that the small body, in fact, could be a BH binary (BHB).  Previous
numerical scatting experiments have shown that SMBHs can tidally capture BHBs
to bound orbits.  Here we investigate the subsequent long-term evolution. We
find that those BHBs with a semi-major axis of $a\lesssim5\times10^{-3}$
AU can be captured to tightly-bound orbits such that they will successfully
inspiral towards the central SMBHs without being scattered away by stellar
relaxation processes.  We estimate that these binary-EMRIs (b-EMRIs) could
constitute at most $10\%$ of the EMRI population.  Moreover, we show
that when the eccentricity of a b-EMRI drops to about $0.85$, the two stellar
BHs will quickly merge due to the tidal perturbation by the SMBH. The
high-frequency ($\sim10^2$ Hz) GWs generated during the coalescence coincide
with the low-frequency ($\sim10^{-3}$ Hz) waves from the b-EMRI, making this
system an ideal target for future multi-band GW observations.  
\end{abstract}


\maketitle

\section{Introduction}\label{sec:intro}

An extreme-mass-ration inspiral (EMRI) normally consists of a stellar compact object,
such as a stellar-mass black hole (BH), and a supermassive black hole
(SMBH).  It is an important target for space-borne, milli-Hz gravitational-wave
(GW) detectors  because it could dwell in the band for as long as the lifetime of
the detectors \citep[e.g.][]{amaro-seoane07,babak17}. Consequently, the number
of GW cycles accumulated in the band approaches $10^{3}-10^{4}$. Such a long
waveform contains rich information about the space-time, as well as the
astrophysical environment, at the immediate exterior of a SMBH
\citep{gair13,barausse14}. To decode this information, our model of an EMRI has
to be exquisitely accurate, both mathematically and physically.

In the canonical model of an EMRI, the stellar object is captured by the SMBH in two
possible ways \citep[see][for a review]{amaro-seoane07}. (i) It is scattered by
other stars, a stochastic process know as relaxation, to such a small distance
to the SMBH that the stellar object loses a significant amount of its orbital
energy through GW radiation and becomes bound to the big BH
\citep{hils95,sigurdsson97}. The event rate of this type of EMRIs is difficult
to estimate because it depends on factors that are poorly constrained by
observations. The current estimation lies in a broad range between $10^{-9}$
and $10^{-6}$ per galaxy per year
\citep{freitag01,hopman05,hopman09,pau11,aharon16,bar-or16,babak17}.  (ii) The
small body can also come from a binary, and now the binary is scattered to the
vicinity of the SMBH.  If the distance between the SMBH and the center-of-mass
of the binary becomes smaller than the tidal radius, $R_t:=a(M_3/m_{12})^{1/3}$
where $a$ and $m_{12}$ are the semi-major axis and total mass of the binary and
$M_3$ the mass of the SMBH, the interaction in general ejects the lighter
member of the binary and leaves the other, more massive member on a bound orbit
around the SMBH \citep{hills88,hills91,miller05}. The corresponding event rate
could be comparable to the previous one \citep{miller05}. 

In this Letter we point to a third possibility: A SMBH could tidally capture a
BH binary (BHB) to such a bound orbit that the binary, as a single unit,
inspirals towards the SMBH through radiation of GWs. Our work is motivated by
the earlier numerical scattering experiments which show that a binary after a
close encounter with a massive body can gain energy and expand its internal semi-major
axis at the expense of the orbital energy around the massive one
\citep{hills91}. This result hints that the massive body can capture the binary
if the latter loses a substantial amount of its orbital energy. More recent
numerical simulations have confirmed this postulation and explicitly showed
that about $(40-50)\%$ of the binaries scattered to a distance of $(1-5)\,R_t$
become bound to the SMBH \citep{addison15}.

However, it is unclear whether these binaries would remain bound to the SMBH for
a long time, a
necessary condition to form a successful EMRI, or they would be scattered away
by the stars surrounding the SMBH.  Here we address this issue and derive a
criterion for successful inspirals. We refer to those events satisfying our
criterion as the ``binary extreme-mass-ratio inspirals'' (b-EMRIs).

Although the long-term interaction between SMBHs and binaries have been studied
previously
\citep{antonini12,mandel15,prodan15,stephan16,VL16,liu17,petrovich17,bradnick17,hoang17},
these earlier works focus on the binaries that are far away from the SMBHs, so
that the GW radiation due to the orbital motion of the binaries around the
SMBHs is not important to the overall dynamics.  In our problem, however, the
BHBs are much closer so that the GW radiation cannot be neglected.  This is the
key difference of our problem from those previous ones. 

\section{Formation of a b-EMRI}

We consider a general case in which a binary of stellar BHs falls towards a
SMBH along a parabolic orbit.  For tidal capture to happen, the BHB should pass
by the SMBH with a pericenter distance of

\begin{align}
R_p&\sim R_t:= a
\left(\frac{M_3}{m_{12}}\right)^{1/3}\simeq37\,R_g
\left(\frac{a}{10^5\,r_g}\right)\nonumber\\
&\times\left(\frac{2}{1+q}\right)^{1/3}
\left(\frac{m_1}{10\,M_\odot}\right)^{2/3}
\left(\frac{M_3}{10^6\,M_\odot}\right)^{-2/3},\label{eq:Rp}
\end{align}

\noindent
where $R_g=GM_3/c^2$ is the gravitational radius of the SMBH, $c$ is the speed
of light, $m_1$ and $m_2$ are the masses of the two stellar BHs, $q:=m_2/m_1$
is the mass ratio assuming that $m_1\ge m_2$, and  $r_g=Gm_1/c^2$ is the
gravitational radius of the bigger stellar BH.  In the following we
focus on the binaries with $q\simeq1$, because they are likely to form in the
star clusters surrounding SMBHs \citep{ligo16astro,amaro-seoane16,oleary16}.
We also scale $a$ with $10^5\,r_g$ and the reason will become clear later in
this section.  The lifetime of the BHB is determined
by the GW radiation timescale

\begin{align}
t_{\rm gw}&:=\frac{a}{|\dot{a}|}=\frac{5a^4F(e)}{64c\,r_g^3q(1+q)}\label{eq:tgw}\\
&\simeq\frac{1.2\times10^7}{q(1+q)}\left(\frac{m_1}{10\,M_\odot}\right)
\left(\frac{a}{10^5\,r_g}\right)^{4}F(e)
\,{\rm yr}, \label{eq:tgw1}
\end{align}

\noindent
where $\dot{a}$ is the decay rate of the semi-major axis due to GW radiation,
$e$ is the orbital eccentricity, and
$F(e)=(1-e^2)^{7/2}(1+73e^2/24+37e^4/96)^{-1}$ \citep{peters64}. 

At the periapsis passage, because of the tidal interaction with the SMBH, the
binary has a $(40-50)\%$ chance of gaining an energy \citep{addison15}.  We can
quantify this energy gain with $\eta Gm_1m_2/(2a)$, where the efficiency $\eta$
is typically $10\%$ \citep{addison15}. According to energy conservation, the
orbit around the SMBH loses the same amount of energy so that the binary
becomes gravitationally bound to the SMBH. The binding energy is
$E_3\simeq \eta Gm_1m_2/(2a)$.

We now look at the properties of this bound orbit.  From $E_3$ we derive a
semi-major axis of $R\simeq (a/\eta)(M_3/\mu)$. On the other hand, the
pericenter remains to be $R_p$ because of the conservation of angular momentum.
Consequently, we can derive an eccentricity $e_3$ from

\begin{align}
1-e_3&=\frac{R_p}{R}\simeq4.6\times10^{-5}\,
\frac{q}{(1+q)^{4/3}}
\left(\frac{\eta}{0.1}\right)\nonumber\\
&\times\left(\frac{m_{1}}{10\,M_\odot}\right)^{2/3}
\left(\frac{M_3}{10^6\,M_\odot}\right)^{-2/3}.\label{eq:e3}
\end{align}

\noindent
Interestingly, it does not depend on $a$. Therefore, the small value of $1-e_3$ indicates
that the orbit of a captured binary, in general, is very eccentric.

The high eccentricity will affect the stability of the orbit in two ways.  On
one hand, the orbit is more susceptible to perturbations by the surrounding
stars because the angular momentum, $\sqrt{GM_3R(1-e_3^2)}$, is small. To be
more quantitative, suppose  $T_{\rm rlx}$ characterizes the typical timescale
for stellar relaxation processes to completely alter the orbital elements of a
circular orbit, the timescale to spoil an orbit with an eccentricity of $e_3$
is only $T_{\rm rlx}(1-e_3^2)$ \citep{amaro-seoane07}. On the other hand, the
orbit circularizes very fast due to GW radiation because the associated
timescale $T_{\rm gw}$ (on which $1-e_3$ decreases) is proportional to
$R^4(1-e_3^2)^{7/2}$ \citep{peters64}. For our purpose, we use the relationship
$R_p=R(1-e_3)$  to rewrite $T_{\rm gw}$ and find that

\begin{align}
T_{\rm gw}&\simeq\frac{13R_p^4}{64 c R_g^3}\left(\frac{M_3}{m_{12}}\right)(1-e_3)^{-1/2}\\
&\simeq4.7\times10^6\,(1+q)^{-7/3}
\left(\frac{1-e_3}{10^{-5}}\right)^{-1/2}\nonumber\\
&\times
\left(\frac{m_1}{10\,M_\odot}\right)^{5/3}
\left(\frac{M_3}{10^6\,M_\odot}\right)^{-2/3}
\left(\frac{a}{10^5\,r_g}\right)^{4}{\rm yr}.\label{eq:Tgw1}
\end{align}

To become a successful b-EMRI, the BHB should be able to circularize.  This
criterion means $(1-e_3^2)\, T_{\rm rlx}>T_{\rm gw}$.  Together with
Equations~(\ref{eq:tgw1}) and (\ref{eq:e3}), we find that the criterion is
equivalent to

\begin{align}
a<a_{\rm cri}&\simeq4.5\times10^4\,r_g\,
\left(\frac{T_{\rm rlx}}{10^9\,{\rm yr}}\right)^{1/4}
\frac{q^{3/8}}{(1+q)^{-1/12}}
\nonumber\\
&\times
\left(\frac{\eta}{0.1}\right)^{3/8}
\left(\frac{m_1}{10\,M_\odot}\right)^{-1/6}
\left(\frac{M_3}{10^6\,M_\odot}\right)^{-1/12}{\rm yr},
\end{align}

\noindent 
where we have adopted a typical value of $10^9$ years for $T_{\rm rlx}$.  This
is the reason that we scaled $a$ with $10^5\,r_g$ in the previous equations.

To estimate the formation rate of b-EMRIs, we first recall that in a relaxed
stellar system it takes approximately a timescale of $T_{\rm rlx}$ for a star
to explore all possible orbital angular momentum around a SMBH, and the
fraction of the orbits that have been explored scales linearly with time
\citep{bt08}. For this reason, during the lifetime of a BHB with $a\sim a_{\rm
cri}$, the probability of being captured is

\begin{align}
p\sim\frac{t_{\rm gw}(a_{\rm cri})}{T_{\rm rlx}}&\simeq4.9\times10^{-4}\,
\frac{q^{1/2}F(e)}{(1+q)^{2/3}}
\left(\frac{\eta}{0.1}\right)^{3/2}\nonumber\\
&\times
\left(\frac{m_1}{10\,M_\odot}\right)^{1/3}
\left(\frac{M_3}{10^6\,M_\odot}\right)^{-1/3}.
\end{align}

\noindent
It turns out that this probability does not depend on our assumption of $T_{\rm
rlx}$.  Furthermore, the typical merger rate of BHBs in a galactic nucleus is
about $\Gamma_{\rm BHB}\sim{\rm few}\,\times10^{-7}$ per year
\citep{miller09VL}.  From these numbers, we deduce a b-EMRI rate of
$p\Gamma_{\rm BHB}\sim 10^{-10}\,{\rm yr^{-1}\,galaxy^{-1}}$. Comparing this
number with the event rate of normal EMRIs, i.e., those consist of single
stellar BHs, which is about $10^{-9}-10^{-6}\,{\rm yr^{-1}\,galaxy^{-1}}$ (see
Section~\ref{sec:intro}), we find that in the most optimistic case b-EMRIs
could constitute $10\%$ of the total EMRI population.

\section{Circularization}

The tidal-capture process described in the previous section produces a triple
system: A BHB revolves around a SMBH on a gravitationally bound and extremely
eccentric orbit. It satisfies the criterion of stability,

\begin{equation}
\frac{a}{R_p}\frac{e_3}{1+e_3}<0.1
\end{equation}

\noindent
\citep{naoz16}. Therefore, we can treat it as a hierarchical triple and study
its secular evolution by dividing it into two components: (i) an ``inner
binary'', which is simply the BHB, and (ii) an ``outer binary'' whose first
member is the BHB, behaving as a single unit, and the second one is the SMBH.

Unlike other triple systems that have been studied previously, our b-EMRI loses
energy and angular momentum due to GW radiation. Moreover, both the inner and
outer binaries emit GWs and are shrinking. It is, therefore, useful to
understand which binary evolves faster.  The evolution timescales have been
calculated in Equations~(\ref{eq:tgw1}) and (\ref{eq:Tgw1}), and from them we
find

\begin{align}
\frac{t_{\rm gw}}{T_{\rm gw}}
&\simeq6.6\,F(e)\,q^{-1}\left(\frac{1+q}{2}\right)^{4/3}\,
\left(\frac{1-e_3}{10^{-5}}\right)^{1/2}\nonumber\\
&\times
\left(\frac{m_1}{10\,M_\odot}\right)^{-2/3}
\left(\frac{M_3}{10^6\,M_\odot}\right)^{2/3}.\label{eq:tgwTgw}
\end{align}

\noindent
Since the majority of the captured binaries have $e<0.5$ \citep{addison15}, the
above ratio is bigger than unity. This result implies that the outer binary
normally evolves faster. In other words, the b-EMRI circularizes first. 

During the process of circularization, the periapsis of the outer binary is
more or less conserved, but the apoapsis keeps shrinking
\citep[e.g.][]{peters64}. As a result, the inner binary feels an increasingly
strong tidal force from the SMBH. The tidal perturbation might excite the
eccentricity of the inner binary due to a Lidov-Kozai mechanism
\citep{lidov62,kozai62,naoz16} and significantly shorten its lifetime $t_{\rm
gw}$. The remaining of this section shows that it happens at a much later time,
not until the outer binary has significantly circularized. 

A necessary condition for the Lidov-Kozai mechanism to take effect is that the
associated timescale, $t_{\rm LK}$, is shorter than the relativistic precession
timescales of the inner and outer binaries
\citep[e.g.][]{chen11,chen13,chen14}.  The precession timescales can be
calculated with $t_{\rm GR}=a(1-e^2)P_{12}/(6\pi r_g)$ and $T_{\rm
GR}=R(1-e_3^2)P_{3}/(6\pi R_g)$, where $P_{12}$ and $P_3$ are, respectively,
the orbital periods of the inner and outer binaries. On the other hand, the
Kozai-Lidov timescale is

\begin{align}
t_{\rm LK}&\simeq\frac{m_{12}}{M_3}\left(\frac{R_p}{a}\right)^{3}
\frac{P_{12}}{(1-e_3)^{3/2}}\simeq\frac{P_{12}}{(1-e_3)^{3/2}}\label{eq:tLK}\\
&\simeq1.0\times10^4\,(1+q)^{-1/2}
\left(\frac{1-e_3}{10^{-5}}\right)^{-3/2}
\nonumber\\
&\times
\left(\frac{m_1}{10\,M_\odot}\right)
\left(\frac{a}{10^5\,r_g}\right)^{3/2}
{\rm yr}
\end{align}
 
\noindent
\citep[e.g.][]{naoz16}, where in Equation~(\ref{eq:tLK}) we have replaced 
$R_p$ with Equation~(\ref{eq:Rp}).  From the above timescales we find
that

\begin{align}
\frac{T_{\rm GR}}{t_{\rm LK}}&\simeq(1+q)^{1/2}\left(\frac{R_p}{3\pi R_g}\right)>1,\\
\frac{t_{\rm GR}}{t_{\rm LK}}&=(1-e^2)(1-e_3)^{3/2}
\left(\frac{a}{6\pi r_g}\right)\\
&\simeq1.7\times10^{-4}(1-e^2)
\left(\frac{1-e_3}{10^{-5}}\right)^{3/2}
\left(\frac{a}{10^5r_g}\right).\label{eq:tGRtLK}
\end{align}

\begin{figure}[!h] 
\begin{center}
\includegraphics[height=2.5in]{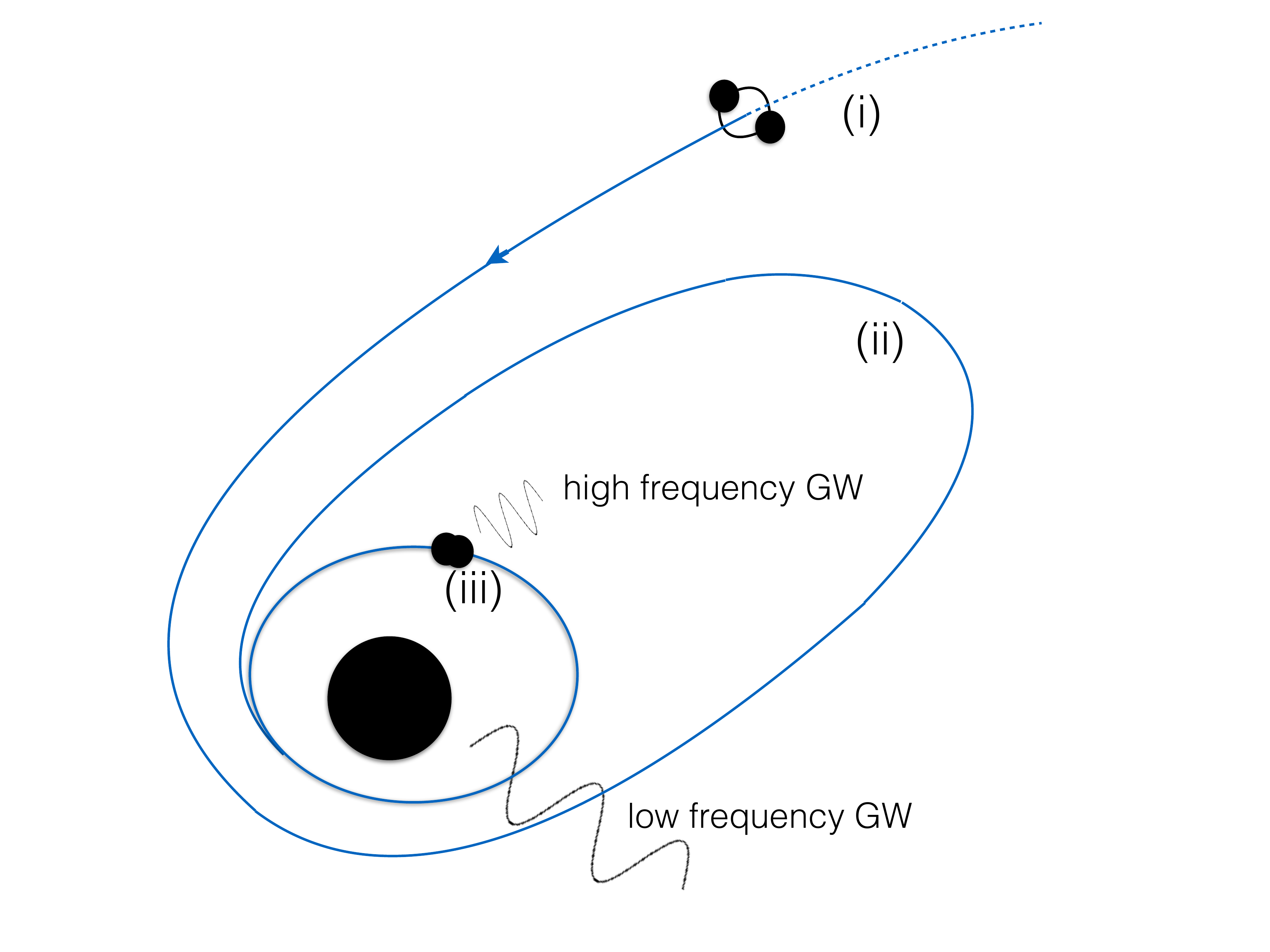}
\caption{Three evolutionary stages of a b-EMRI. (i) A compact BHB is captured
to a bound orbit around a SMBH because the pericenter distance becomes comparable
to the tidal-disruption radius, $R_t$. (ii) The outer binary circularizes due
to GW radiation, and the GW frequency lies in the band of a space-borne GW
detector. (iii) When the eccentricity of the outer binary decreases to about
$0.85$, the tidal force of the SMBH becomes strong enough to excite the
eccentricity of the inner BHB and drive it to merge. High-frequency GWs
detectable by ground-based observatories are emitted during the merger,
together with the low-frequency waves from the outer binary.  }
\label{fig:bEMRI} \end{center} \end{figure}

Now that $T_{\rm GR}>t_{\rm LK}$, the precession of the outer binary does not
quench the Lidov-Kozai mechanism. However, the precession of the inner binary
does, because $t_{\rm GR}\lesssim t_{\rm LK}$ unless $1-e_3\gtrsim10^{-2}$.
Therefore, the outer binary circularizes first and the merger of the inner BHB
happens at a much later time.  Figure~\ref{fig:bEMRI} illustrates this general
picture.

\section{Aborting a b-EMRI}

We now study the late evolution of the b-EMRI, i.e. after $1-e_3$ has increased
to about $0.01$.  We have seen from the last section that the Lidov-Kozai
mechanism starts to affect the evolution of the inner BHB. Since we find that 

\begin{align}
\frac{t_{\rm LK}}{t_{\rm gw}}&\simeq2.6\times10^{-8}\,
\frac{q(1+q)^{1/2}}{F(e)}
\left(\frac{1-e_3}{10^{-2}}\right)^{-3/2}
\left(\frac{a}{10^5r_g}\right)^{-5/2},\label{eq:tgwtLK}\\
\frac{t_{\rm LK}}{T_{\rm gw}}&\simeq2.1\times10^{-6}\,(1+q)^{11/6}
\left(\frac{1-e_3}{10^{-2}}\right)^{-1}
\nonumber\\
&\times
\left(\frac{m_1}{10\,M_\odot}\right)^{-2/3}
\left(\frac{M_3}{10^6\,M_\odot}\right)^{2/3}
\left(\frac{a}{10^5\,r_g}\right)^{-5/2},
\end{align}

\noindent
GW radiation is not important on the Lidov-Kozai timescale, at least at the
beginning of the interaction when $e\lesssim0.5$.  Therefore, both the energy
and angular momentum of the triple is conserved, so that we can use the
standard Lidov-Kozai scenario to predict the subsequent evolution.

The Lidov-Kozai mechanism in general drives the eccentricity of the inner
binary, $e$, to evolve between a maximum and a minimum value determined by the
initial conditions of the triple.  Consequently, $t_{\rm gw}$ varies according
to the function of $F(e)$.  If the maximum $e$ is so high that $t_{\rm gw}$
becomes shorter than $t_{\rm LK}$, the inner BHB decouples from the triple and
merges.  The merger terminates the b-EMRI and transforms it into a standard
EMRI. 

To find out when the decoupling will happen, we re-evaluate
Equation~(\ref{eq:tgwtLK}).  The small numerical coefficient indicates that
only when $F(e)\lesssim10^{-7}$ can $t_{\rm gw}$ be comparable to or shorter
than $t_{\rm LK}$, i.e. $e$ is extremely large when the decoupling happens.  On
the other hand, $e$ cannot exceed the limit imposed by the condition $t_{\rm
GR}>t_{\rm LK}$. Otherwise, 
the relativistic precession will quench
the Lidov-Kozai mechanism.

Therefore, terminating a b-EMRI requires that $t_{\rm gw}<t_{\rm LK}<t_{\rm
GR}$. By evaluating Equations~(\ref{eq:tGRtLK}) and (\ref{eq:tgwtLK}), we find
that the necessary condition for the binary to merge is

\begin{equation}
1-e_3\gtrsim0.15\,q^{-4/15}(1+q)^{-2/15}\left(\frac{a}{10^5\,r_g}\right)^{-4/15}.\label{eq:e3cri}
\end{equation}

\noindent
This condition does not explicitly depend on mass, implying that the merger
will happen regardless of the composition of the binary (be it a BH or a
neutron-star binary).

To prove that the above condition is sufficient, we run numerical simulations
to study the evolution of $e$.  Our method is different from those used in the
earlier studies of a BHB around a SMBH.  We cannot adopt the standard
scheme of ``double-average'' \citep{naoz16} to integrate our triple system
because the evolutionary timescale $t_{\rm LK}$ is comparable to the orbital
period of the outer binary $P_3$. One can see this by substituting the term
$1-e_3$ in Equation~(\ref{eq:tLK}) with $1-e_3=R_p/R$.  For this reason, we use
the scheme developed in \citet{luo16LK} to orbital-average only the inner
binary and numerically solve their Equation (19) to get the evolution of $e$.
Although this ``single-average'' scheme does not
account for the effect of relativistic precession, this caveat would not
significantly change our conclusion as long as $e_3\lesssim0.85$, as we have
shown by deriving Equation~(\ref{eq:e3cri}).

\begin{figure}
\centering
\includegraphics[width=0.5\textwidth]{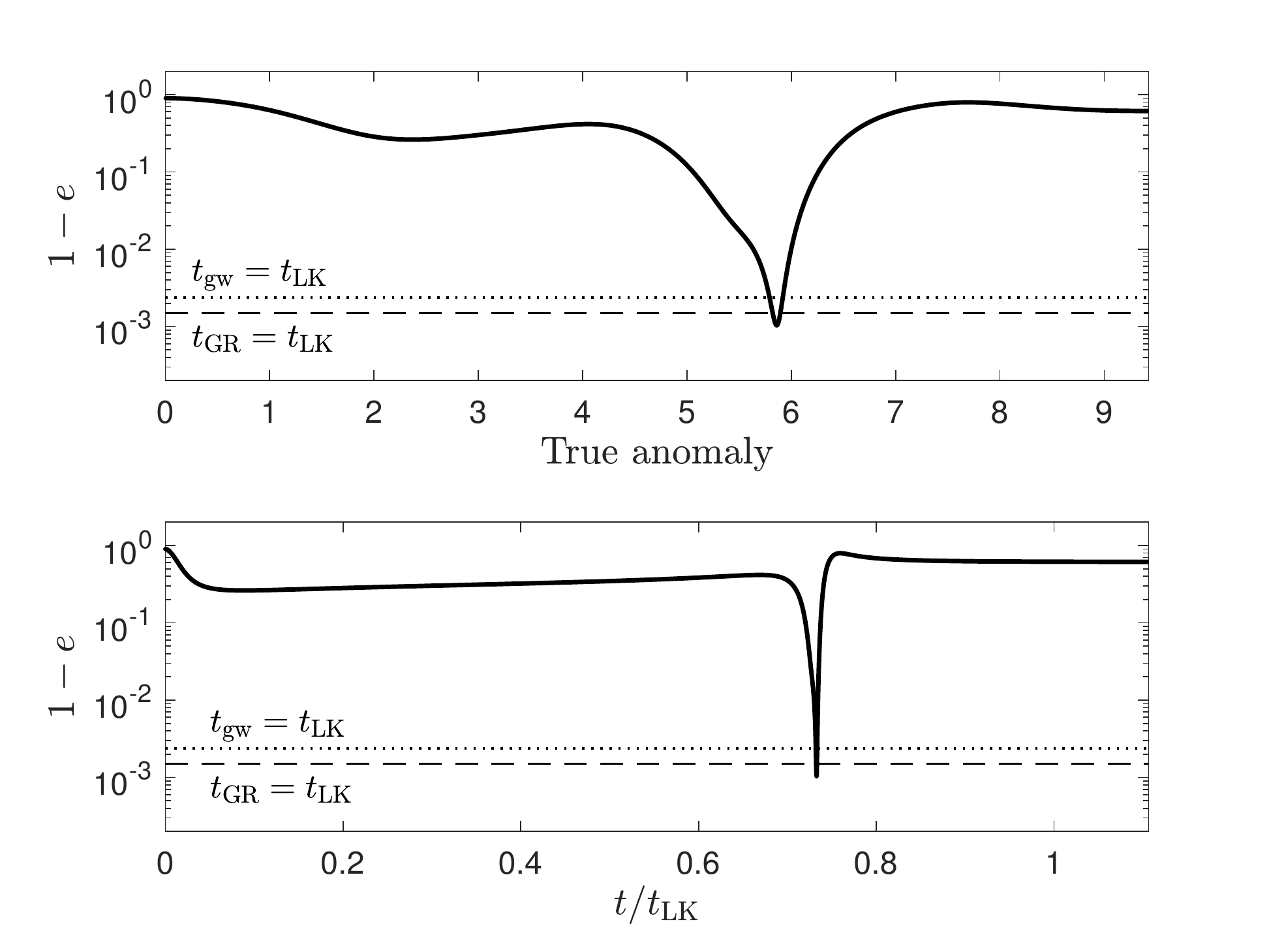}
\caption{Excitation of the eccentricity of a BHB during its final orbit around
the SMBH. The upper (lower) panel shows the evolutionary track (solid line) as
a function of the true anomaly (time).  Here we choose $m_1=m_2=10\,M_\odot$,
$M_3=10^6\,M_\odot$, $a=5\times10^4\,r_g$, $R_p=1.4R_t$, $e=0.1$, and
$e_3=0.75$ as the initial conditions. The inclination
angle between the inner and outer binaries initially is $10\,\deg$.
 The BHB starts at the periapsis where the
true anomaly is $0$. When it completes an orbit around the SMBH and returns to
the periapsis, the eccentricity is excited to such a large value that the
condition for a merger, $t_{\rm gw}=t_{\rm LK}$ (dotted lines), is satisfied.
The condition for the relativistic precession to quench the Lidov-Kozai
mechanism, i.e. $t_{\rm GR}=t_{\rm LK}$, is indicated by the dashed lines.}
\label{fig:ecc075} \end{figure}

Figure~\ref{fig:ecc075} shows an example of our numerical simulation.  It
confirms our prediction that $e$ is excited to a large value such that the
condition for a merger, $t_{\rm gw}=t_{\rm LK}$, can be satisfied.  This
happens close to the periapsis passage, where the true anomaly is $2\pi$.  The
figure also suggests that the relativistic precession would not quench the
Kozai-Lidov mechanism because an even larger $e$ is needed.

\section{Discussions}

In this Letter we have presented a new type of EMRIs. They form due to tidal
capture of BHBs by SMBHs. We refer to them as b-EMRIs. 

We find that the binaries should meet the criterion of $a\lesssim a_{\rm
cri}\simeq 5\times10^4\,r_g$ to overcome the perturbations by the surrounding
stars and inspiral successfully towards the SMBHs. Interestingly, these
binaries are detectable by a space-borne GW observatory, such as the Laser
Interferometer Space Antenna \citep[LISA,][]{lisa17}.  This is so because LISA
is sensitive to the GWs with a frequency of $f\sim10^{-3}$ Hz and the strongest
GW mode that a BHB emit is of a frequency of
$f\simeq\pi^{-1}\sqrt{Gm_{12}/[a(1-e)]^3}$ \citep{farmer03,wen03}. As a result,
those BHBs with a semi-major axis of

\begin{equation}
a\sim3.4\times10^4\,r_g\,
f_{-3}^{-2/3}
\frac{(1+q)^{1/3}}{1-e}
\left(\frac{m_1}{10\,M_\odot}\right)^{-2/3}\label{eq:a_lisa}
\end{equation}

\noindent
are inside the LISA band, where  $f_{-3}:=f/(10^{-3}\,{\rm Hz})$.

On the other hand, the orbital motion of the BHBs around the SMBHs also
generate GWs. It is important to understand whether this radiation is also
detectable.  Similar to the previous analysis for BHBs, we calculate the
frequency of the strongest GW mode as $f=\pi^{-1}\sqrt{GM_3/R_p^3}$, so that a
b-EMRI is inside the LISA band if its periapsis is 

\begin{equation}
R_p\sim16\,R_g\,
f_{-3}^{-2/3}
\,M_6^{-2/3}.\label{eq:R_lisa}
\end{equation}

\noindent
We notice that the above criterion $a\lesssim a_{\rm cri}$ indicates that
our b-EMRIs typically have

\begin{align}
R_p&\lesssim21\,R_g\,
\left(\frac{T_{\rm rlx}}{10^9\,{\rm yr}}\right)^{1/4}
\frac{q^{3/8}}{(1+q)^{1/4}}
\nonumber\\
&\times
\left(\frac{\eta}{0.1}\right)^{3/8}
\left(\frac{m_1}{10\,M_\odot}\right)^{1/2}
\left(\frac{M_3}{10^6\,M_\odot}\right)^{-3/4}.
\end{align}

\noindent
Therefore, they are indeed detectable by LISA. Now we have a source
which generates two types of GWs, i.e from the BHB and its orbit around the
SMBH, in the same band and at the same time. It is extremely interesting from
the observational point of view. In particular, the phase of the GWs from the
BHB will be modulated by the orbital motion around the SMBH
\citep{inayoshi16,meiron17}. 

Moreover, we have shown that GW radiation, in general, circularizes a b-EMRI.
However, when the eccentricity of the outer binary has decreased to about
$e_3\sim0.85$, the tidal force of the SMBH is strong enough to excite the
eccentricity of the BHB to an extreme value and trigger the merger of the
binary. As a result, the b-EMRI aborts before the BHB gets too close to the
SMBH.

The termination of the b-EMRI provides a rare but valuable case for future
multi-band GW observations.  First, the merger generates high-frequency GWs,
i.e. a LIGO/Virgo event that is simultaneous with and in the same sky location
of a LISA EMRI event. Second, the high-frequency GWs could be redshifted
because they are generated very close to a SMBH \citep{chen17}, providing a
rare opportunity of studying the propagation of GWs in the regime of strong
gravity.  Third, the merger also induces a kick to the BH remnant
\citep{centrella10}.  This kick causes a glitch in the EMRI waveform, which,
through a careful analysis, is discernible in the data stream \citep{han18}.

\acknowledgements

We thank Subo Dong and Fukun Liu for many discussions.  This work is supported
by NSFC No.  U1431120, No.11273045, the ``985 Project'' of Peking University,
and partly by the Strategic Priority Research Program of the Chinese Academy of
Sciences, Grant No. XDB23040100 and No. XDB23010200.  The authors also thank
Pau Amaro-Seoane and Carlos Sopuerta for organizing the 2017 Astro-GR
Meeting@Barcelona, where the idea of this work was conceived.

\bibliographystyle{astroads.bst}

\end{document}